\documentclass[aps,prl,twocolumn,showpacs,superscriptaddress,groupedaddress]{revtex4-1}
\pdfoutput=1
\usepackage[usenames,dvipsnames]{xcolor}  %%%%%%% usenames,dvipsnames required to get BrickRed
\usepackage{colortbl} 
\usepackage{graphicx} %%%%%% do not use \usepackage[dvips]{graphicx}%%%% it will creat a clash with color n brickred will not appear%%%%%
\definecolor{lightblue}{rgb}{0.93, 0.95, 1.0} % Azul claro
\definecolor{lightgray}{gray}{0.9} % Gris claro
\usepackage{bbold}
\usepackage{amsthm}
\usepackage{amsmath}
\usepackage{slashed}
\usepackage{amssymb}
\usepackage{cancel}
\usepackage{multirow}%\usepackage{dsfont}
\usepackage{hhline}
\usepackage{float}
\usepackage[utf8]{inputenc}
\usepackage{orcidlink}
\usepackage{booktabs}
\usepackage{ulem}
\definecolor{darkred}{rgb}{0.6,0,0}
\usepackage[]{hyperref}

\usepackage{braket}
\definecolor{linkcolor}{rgb}{0,0,0.5}
\allowdisplaybreaks
\allowdisplaybreaks[1]
\allowdisplaybreaks[2]

\def\vev#1{\left\langle #1\right\rangle}

\definecolor{greenLinks}{rgb}{0, 0.6, 0}
\definecolor{blueLinks}{rgb}{0, 0, 0.6}
\definecolor{redLinks}{rgb}{0.6, 0, 0}
\definecolor{tempText}{rgb}{0.55, 0.10,0.67}
\definecolor{eprintLinks}{rgb}{0.4, 0.4, 0.4}
\definecolor{journalLinks}{rgb}{0.6, 0, 0}
\newcommand {\ignore}[1]{}

\usepackage{soul}
\definecolor{mightnightblue}{RGB}{25,25,112}

\definecolor{brown}{rgb}{0.59, 0.29, 0.0}

\def\vev#1{\left\langle #1\right\rangle}

\def\lsim{\mathrel{\rlap{\lower4pt\hbox{\hskip1pt$\sim$}}
    \raise1pt\hbox{$<$}}}
\def\gsim{\mathrel{\rlap{\lower4pt\hbox{\hskip1pt$\sim$}}
    \raise1pt\hbox{$>$}}}

\def\U1s{$\mathrm{U_{1}^{(a)}\otimes U_{1}^{(b)}\otimes U_{1}^{(c)}\otimes U_{1}^{(d)}\otimes U_{1}^{(e)}}$ }
\def\3211{$\mathrm{SU(3) \times SU(2)_L \times U(1)_R \times U(1)_{B-L}}$ }
\def\321{$\mathrm{SU(3) \times SU(2) \times U(1)}$ }
\def\422{$\mathrm{SU(4) \times SU(2) \times SU(2)_R}$ }

 \newcommand{\AddrIFIC}{%
  Instituto de F\'{i}sica Corpuscular, CSIC-Universitat de Val\`{e}ncia, 46980 Paterna, Spain\vspace{0.1cm}}

\newcommand{\AddrBasel}{Department of Physics, University of Basel, Klingelbergstrasse 82, CH-4056 Basel,Switzerland}

\begin{document}

\title{\boldmath \color{BrickRed} Gravity-assisted neutrino masses}

\author{Stefan Antusch\orcidlink{0000-0001-6120-9279}}\email{stefan.antusch@unibas.ch} \affiliation{\AddrBasel}
\author{Salvador Centelles Chuli\'{a}\orcidlink{0000-0001-6579-1067}}\email{salcen@ific.uv.es}
\affiliation{\AddrIFIC}
\author{Miguel Levy\orcidlink{0000-0002-5854-8939}}\email{miguelpissarra.levy@unibas.ch}
\affiliation{\AddrBasel}

%%%%%%%%%%%%%%%%%%%%%%%%%%%%%

\begin{abstract}

Gravity is generally expected to violate global symmetries, including lepton number. However, neutrino masses from the Planck-suppressed Weinberg operator are typically too small to account for oscillation data. We propose a new model-building approach to low-scale neutrino mass generation, in which an intermediate spontaneous symmetry-breaking scale generates masses and mixings in the heavy neutral lepton (HNL) sector, while leaving an unbroken residual symmetry $G_{\mathrm{res}}$ that forbids light-neutrino masses. The observed light-neutrino masses then arise because gravity breaks $G_{\mathrm{res}}$ via Planck-suppressed operators, inducing the small lepton-number violation required in low-scale seesaw constructions. The HNLs form pseudo-Dirac pairs, with masses potentially within reach of future colliders and complementary tests in precision searches such as charged lepton flavour violation (cLFV). As an illustration, we present a representative realisation of this class of models and show that, for $\mathcal{O}(1)$ operator coefficients, it predicts a region in the ($M_R$, $\Theta^2$)-plane that can be testable via displaced-vertex searches at the High-Luminosity (HL) LHC and the FCC-ee.

\end{abstract}

%%%%%%%%%%%%%%%%%%%%%%%%%%%%%

\maketitle

%%%%%%%%%%%%%%%%%%%%%%%%%%%%%%%
\section{Introduction}
\label{sec:intro}
%%%%%%%%%%%%%%%%%%%%%%%%%%%%%

The small but non-zero neutrino masses have remained a puzzle ever since the discovery of neutrino oscillations~\cite{Kajita:2016cak, McDonald:2016ixn}. In the Standard Model (SM), neutrinos are massless due to an accidental $U(1)$ global symmetry. However, gravity is expected to violate all global symmetries~\cite{Hawking:1975vcx,Barbieri:1979hc,Kallosh:1995hi,Banks:1988yz,Banks:2010zn,Harlow:2018tng}, inducing \textit{e.g.}~the Planck-suppressed Weinberg operator~\cite{Weinberg:1979sa}
\begin{subequations}
\label{eq:Weinbergdirect}
\begin{align}    
   \mathcal{O}_W^\text{gravity} &= \dfrac{\mathcal{G}_W}{M_P}\,\bar{L}^cHHL 
   \\ 
   \Rightarrow\quad 
   m_\nu^\text{gravity} &= \mathcal{G}_W\dfrac{v^2}{M_P} \sim 10^{-6}\,\text{eV}\, .
\end{align}    
\end{subequations}
Earlier attempts to exploit gravity-induced effects in neutrino masses were explored in \cite{Akhmedov:1992hh,Grasso:1992fv,deGouvea:2000jp}. However, the KamLAND measurement of $\Delta m_{21}^2$ \cite{KamLAND:2002uet} definitively ruled out such scenarios. Clearly, the Planck-suppressed Weinberg operator yields neutrino masses far below the oscillation scales~\cite{deSalas:2020pgw}, $\left(\sqrt{\Delta m_{31}^2}, \sqrt{\Delta m_{21}^2} \right) \approx  (0.05, 0.009)$ eV. Further developments connecting neutrino masses with gravity-induced effects include \cite{Borah:2013mqa,Carvajal:2015dxa,Ibarra:2018dib,Dvali:2016uhn,Barenboim:2019fmj,Borah:2019ldn,Davoudiasl:2020opf,Senjanovic:2020rcq,Borah:2020ljr}.

A well-known mechanism capable of explaining the observed neutrino masses is the standard type-I seesaw~\cite{Minkowski:1977sc,Yanagida:1979as,Mohapatra:1979ia,Gell-Mann:1979vob,Schechter:1980gr}, in which HNLs break $U(1)$ explicitly through their Majorana mass terms. Integrating them out of the theory generates the Weinberg operator suppressed by $M_R$ instead of $M_P$, yielding the correct neutrino masses for $M_R\sim 10^{14}\,\text{GeV}$. Neutrino masses in the standard type-I seesaw mechanism are small because $M_R$ is large. However, in addition to the large masses of the HNLs beyond scope of any envisioned collider, the resulting active--sterile mixing $\Theta$ is of order $\Theta^2 \sim m_\nu/M_R$, rendering indirect tests via precision experiments unobservable. 

In contrast to the standard type-I seesaw, genuine low-scale versions, such as the inverse~\cite{Mohapatra:1986bd,GonzalezGarcia:1988rw} and linear~\cite{Akhmedov:1995ip,Akhmedov:1995vm,Malinsky:2005bi,Antusch:2017tud} seesaws, allow for lower HNL masses and sizeable active--sterile mixing leading to a rich phenomenology. Representative examples include cLFV~\cite{Bernabeu:1987gr,Gonzalez-Garcia:1988okv,Ilakovac:1994kj,Fujihara:2005uq,Antusch:2006vwa,Abada:2014kba,Lindner:2016bgg,Hagedorn:2021ldq,Herrero-Brocal:2023czw,CentellesChulia:2024uzv,CentellesChulia:2025eck}, 
collider signatures~\cite{Dittmar:1989yg,Gonzalez-Garcia:1990sbd,delAguila:2007qnc,Atre:2009rg,Antusch:2020pnn,Aguilar-Saavedra:2012dga,Das:2012ze,Das:2012ii,Deppisch:2013cya,Antusch:2015mia,Deppisch:2015qwa,Antusch:2016vyf,Antusch:2016qby,Antusch:2016ejd,Antusch:2017hhu,Antusch:2017ebe,Antusch:2017pkq,Bondarenko:2018ptm,Antusch:2018bgr,Abada:2018sfh,Antusch:2019eiz,Hirsch:2020klk,Antusch:2022ceb,Cottin:2022nwp,Antusch:2022hhh,Chauhan:2023faf,Antusch:2023nqd,Batra:2023mds,Antusch:2023nqd,Batra:2023ssq,Antusch:2024otj}, 
and non-standard neutrino interactions~\cite{Antusch:2008tz,Antusch:2014woa,Escrihuela:2015wra,Blennow:2016jkn}. 
These constructions rely on the presence of a small lepton-number--violating parameter $\mu$, where taking $\mu$ to zero results in vanishing light neutrino masses.  
However, although technically natural~\cite{tHooft:1979rat}, a small $\mu$ is still introduced \textit{ad hoc}. This leads to the question of the model building framework behind such scenarios.

We pursue the following strategy. We consider an intermediate sector at a scale $\Lambda$ with $M_P \gg \Lambda \gg \Lambda_{\rm EW}$, together with a global symmetry $G_\Lambda$. We introduce a set of neutral leptons, $N_i$, the would-be seesaw mediators, whose masses and mixings with the active neutrinos are forbidden by $G_\Lambda$.

A complex scalar $\phi$ acquires a vacuum expectation value $\langle \phi \rangle < \Lambda$, spontaneously breaking $G_\Lambda$ to a residual symmetry $G_{\rm res}$. Via the mediation of fields at the scale $\Lambda$, this breaking generates masses for the HNLs $N_i$, as well as their mixing with the active neutrinos through neutrino Yukawa interactions.

The residual subgroup $G_{\rm res}$ continues to protect the light-neutrino masses, implying $m_\nu = 0$ at this stage, in the absence of gravity-induced effects.

Light neutrino masses finally arise when gravity breaks $G_{\rm res}$ through Planck-suppressed operators, inducing a low-scale seesaw. The interplay of these two symmetry breakings allows for the required enhancement of the Planck-suppressed Weinberg operator in Eq.~\eqref{eq:Weinbergdirect} to account for oscillation data.

In a simple realisation with two HNLs, the UV symmetry is identified with $G_\Lambda=U(1)$, which is spontaneously broken by a scalar $\phi$ with charge $q(\phi)=3$, leaving a residual $Z_3$ symmetry which protects both the Dirac nature of the HNLs $N_1$ and $N_2$, as well as the vanishing of light-neutrino masses. Gravity breaks the residual $Z_3$ symmetry, inducing a small lepton-number violation. In what follows, we detail the theoretical framework of this minimal realisation, which represents the simplest example of a broader class of gravity-assisted low-scale seesaws, where both the neutrino mass scale and the HNL phenomenology are controlled by the symmetry-breaking scale in the intermediate sector.

In this novel approach, Planck-induced lepton-number violation is the sole source of neutrino masses and is parametrically enhanced by an intermediate symmetry-breaking sector, leading to predictive and testable low-scale seesaw realisations.

%%%%%%%%%%%%%%%%%%%%%%%%%%%%%%
\section{Theoretical framework}
\label{sec:setup}
%%%%%%%%%%%%%%%%%%%%%%%%%%%%%%

Following the strategy outlined above, we extend the SM by a global $U(1)$ symmetry, a complex scalar $\phi$ with charge $q(\phi)=3$, and a pair of heavy neutral leptons $N_1$ and $N_2$ with charges $q(N_1,N_2)=(1-3n_1,-4+3n_1-3n_2)$, for a given choice of integers $(n_1,n_2)$. The scalar $\phi$ spontaneously breaks $U(1)$ down to a residual $Z_3$, and an intermediate sector at the scale $\Lambda$ is integrated out. The effective Lagrangian and neutral fermion mass matrix read as
\begin{subequations}
\begin{align}
    \mathcal{L} &= \mathcal{C}_1 \, \bar{L} H N_1  \dfrac{\phi^{n_1}}{\Lambda^{n_1}}  +  \mathcal{C}_2\bar{N}_1^c N_2 \, \phi  \dfrac{\phi^{n_2}}{\Lambda^{n_2}}  \, , \\     
    \mathcal{M}^\text{sym}_\nu &= \begin{pmatrix}
        0 & \mathcal{C}_1 \, v \dfrac{\vev{\phi}^{n_1}}{\Lambda^{n_1}} & 0 \\
        . & 0 & \mathcal{C}_2 \vev{\phi} \, \dfrac{\vev{\phi}^{n_2}}{\Lambda^{n_2}}  \\
        . & . & 0
    \end{pmatrix} \, . 
\end{align}
\end{subequations}   
The residual $Z_{3}$ symmetry protects neutrino masses, implying $m_\nu = 0$ in the limit $M_P \to \infty$, irrespective of the choice of $(n_1, n_2)$. A Dirac mass for the HNLs and their mixing with the active neutrinos arise after spontaneous symmetry breaking, while the vanishing entries of $\mathcal{M}_\nu^{\rm sym}$ remain protected.

Including Planck-suppressed symmetry-breaking effects, the effective neutral fermion mass matrix takes the form
\begin{align}
     \mathcal{M}_\nu = 
     \begin{pmatrix}
        \mathcal{G}_\text{W} \dfrac{v^2}{M_P} & Y v \dfrac{\vev{\phi}^{n_1}}{\Lambda^{n_1}} & \mathcal{G}_L v \dfrac{\vev{\phi}}{M_P} \\
        . & \mathcal{G}_\ell \dfrac{\vev{\phi}^2}{M_P} & \mathcal{C}_2 \vev{\phi}  \dfrac{\vev{\phi}^{n_2}}{\Lambda^{n_2}} \\
        . & . & \mathcal{G}_i \dfrac{\vev{\phi}^2}{M_P}
     \end{pmatrix} \, , 
\end{align}
where the gravity-induced corrections to the non-vanishing terms are assumed to be sub-dominant.

Each of the gravity-induced terms gives a contribution to neutrino masses. $\mathcal{G}_\text{W}$ induces a direct Weinberg-like term, too small to satisfy oscillation observations, as argued in connection with Eq.~\eqref{eq:Weinbergdirect}. $\mathcal{G}_L$ leads to a Linear Seesaw-type (LS) mass term, while $\mathcal{G}_\ell$ is a radiative contribution~\cite{Dev:2012sg}. Finally, $\mathcal{G}_i$ gives an Inverse Seesaw-type (IS) contribution.

We focus on the one-generation limit to highlight the framework, leaving the flavour structure for future work. The dominant contributions to the light neutrino masses are given by ($\varepsilon \equiv \langle \phi \rangle / \Lambda$):
\begin{subequations}
\label{eq:numasses}    
\begin{align}
    m_\nu^\text{LS} &= 2 \mathcal{G}_L \dfrac{\mathcal{C}_1}{\mathcal{C}_2}\dfrac{v^2}{M_P} \varepsilon^{(n_1-n_2)}\, \quad , \\ 
    m_\nu^\text{IS} &= \mathcal{G}_i \dfrac{\mathcal{C}_1^2}{\mathcal{C}_2^2} \dfrac{v^2}{M_P} \varepsilon^{2(n_1-n_2)}\, \quad . 
\end{align}
\end{subequations}
For the sake of completeness, we also present the contribution from the loop-induced seesaw~\cite{Dev:2012sg}:
\begin{subequations}
\begin{align}\label{eq:loopinducedSeesaw}
    m_\nu^\ell = 
    m_\nu^\text{IS} \, \dfrac{\mathcal{G}_\ell}{\mathcal{G}_i} \,  \dfrac{g^2}{64 \pi^2} f(M_R) \, , 
\end{align}
with 
\begin{align}\label{eq:loopinducedFunction}
    f(M_R) =  \dfrac{x_{hW}^2}{1-x^2_{hR}} \log {\dfrac{1}{x_{hR}^2}} + \dfrac{3 x_{ZW}^2}{1-x_{ZR}^2} \log {\dfrac{1}{x_{ZR}^2}} \, , 
\end{align}
\end{subequations}
where $x_{ab}=m_a/m_b$, $M_R = \mathcal{C}_2 \vev{\phi} \varepsilon^{n_2}$, $m_h$, $m_W$, and $m_Z$ the higgs, $W$- and $Z$-boson masses, and $g$ the weak coupling constant.   
Due to the loop-suppression, the radiative contribution to $m_\nu$ is expected to be sub-leading compared to the inverse seesaw contribution for the naive expectation $\mathcal{G}_i \sim \mathcal{G}_L  \sim \mathcal{G}_\ell \sim \mathcal{O}(1)$. For this reason, we ignore $\mathcal{G}_\ell$ in what follows.  

As a side note, this may be seen as the expectation when all the Planck-scale suppressed effective operators are generated by perturbative gravity. On the other hand, given our ignorance regarding the full quantum theory of gravity to be applied at the Planck scale, it might be that only some of the operators emerge in this way, while others arise only non-perturbatively with some further suppression. In such scenarios, any of the three low-scale seesaw versions could be the dominant one.  
Regardless, in the following we keep to the assumption $\mathcal{G}_i \sim \mathcal{G}_L  \sim \mathcal{G}_\ell \sim \mathcal{O}(1)$.

The simple result of Eq.~\eqref{eq:numasses} perfectly encapsulates how the interplay between the Planck and intermediate scales can boost the light-neutrino mass scale obtained by the Planck-suppressed operators, generating ``gravity-assisted" neutrino masses.  
The case of $n_1=n_2$ coincides with the naive expectation of Eq.~\eqref{eq:Weinbergdirect}. This contribution is too small, but the intermediate scale, via $\varepsilon$, can enhance it to match the oscillation data. For such an enhancement to occur, assuming $\varepsilon<1$, we must have $n_2 > n_1$ (consequently, $\varepsilon^{(n_1-n_2)} > 1$), to compensate for the largeness of $M_P$.  
In this regime, it is also straightforward to see that, without introducing hierarchies in the couplings, the IS contribution will necessarily dominate over the LS:
\begin{align}\label{eq:dom}
    \dfrac{m_\nu^\text{IS}}{m_\nu^\text{LS}} = \dfrac{1}{2} \dfrac{\mathcal{G}_i}{\mathcal{G}_L} \dfrac{\mathcal{C}_1}{\mathcal{C}_2} \varepsilon^{(n_1-n_2)} > 1 \, , \quad \text{for} \quad %\varepsilon^{(n_1-n_2)} > 1 \, , \quad 
    \dfrac{\mathcal{G}_i \, \mathcal{C}_1}{\mathcal{G}_L\, \mathcal{C}_2} \approx 1 \, .
\end{align}

Delving deeper into Eq.~\eqref{eq:numasses}, we see that the discrete choice of $(n_1, n_2)$ will determine different phenomenological consequences for the model at hand.  
As a consistency condition, these constructions hold only as long as the gravity-induced corrections are sub-dominant with respect to the effective, symmetry-allowed, operators (setting all coefficients to one): 
\begin{align}\label{eq:cond1}
\vev{\phi} \ll M_P \varepsilon^{n_2} <     M_P \varepsilon^{n_1} \, .
\end{align}
The second consistency condition relates to the seesaw expansion, such that we must require
\begin{align}\label{eq:cond2}
\vev{\phi} \gg v \varepsilon^{n_1-n_2} \, .
\end{align}
Under these conditions, the gravity-assisted neutrino masses fall into the category of a low-scale seesaw. Indeed, since $\vev{\phi}$ determines both $M_R$ and $\Theta$, this setup leads to a highly predictive scenario. As this scale is simultaneously bounded from above and below by the consistency conditions of Eqs.~\eqref{eq:cond1}~and~\eqref{eq:cond2}, a given choice of $(n_1,n_2)$ corresponds to a correlated and bounded range for $M_R$ and $\Theta$.

A more general study of this broad class of models and its rich phenomenology is left for a future work. For the sake of concreteness, we dedicate the next section to present a relevant example of this construction. 

%%%%%%%%%%%%%%%%%%%%%%%%%%%%%%%%%%%%%%%%%%%%%%%%%%%%%%%%%%%%%%%%%
\section{A benchmark scenario for the FCC-$\text{ee}$}
%%%%%%%%%%%%%%%%%%%%%%%%%%%%%%%%%%%%%%%%%%%%%%%%%%%%%%%%%%%%%%%%%

Let us now consider the case
\begin{align}
    (n_1, n_2) = (5, 7) \, . \label{eq:n1n2FCC}
\end{align}
As argued before, the couplings of $\bar{L} H N_1$ and $\bar{N}_2^c N_1$ are effective operators which appear after integrating out the messenger fields at some scale $\Lambda$. The charge assignment for this scenario is shown in Table~\ref{tab:charges_base}. 

\begin{table}[b!]
\begin{center}
\renewcommand{\arraystretch}{1.35}
\setlength{\tabcolsep}{5pt}
\begin{tabular}{| c | c | c|}
\hline
Fields & $SU(2)_L \otimes U(1)_Y$ & $U(1) \rightarrow Z_3$ \\ 
\hline
$H$ & $ (\mathbf{2}, \frac{1}{2}) $ & $0 \rightarrow 1$  \\ 
$\phi$ & $ (\mathbf{1}, 0) $ & $3 \rightarrow 1$  \\
\hline
$L$ & $(\mathbf{2}, -\frac{1}{2})$ &  $1 \rightarrow \omega$  \\
$N_1$ & $ (\mathbf{1}, 0) $ & $-14 \rightarrow \omega$ \\
$N_2$ & $ (\mathbf{1}, 0) $ & $-10 \rightarrow \omega^2$  \\
\hline
\end{tabular}
\end{center}
\caption{
Electroweak and $U(1)$ charges of the minimal model with $(n_1, n_2) = (5, 7)$. The scalar $\phi$ spontaneously breaks $U(1)$ to a residual $Z_3$, forbidding all Majorana mass terms in the $M_P\to \infty$ limit. Here $\omega=\exp(i\,2\pi/3)$.
\label{tab:charges_base}}
\end{table}

Considering that the inverse seesaw dominates over the others (cf.~Eq.~\eqref{eq:dom}), we only consider this contribution going forward. Following the general prescription for this particular case we obtain 

\begin{align}
     \mathcal{M}_\nu = 
     \begin{pmatrix}
        0 & \mathcal{C}_1 v \, \varepsilon^5 & 0 \\
        . & 0 & \mathcal{C}_2  \vev{\phi} \, \varepsilon^7 \\
        . & . & \mathcal{G}_i \dfrac{\vev{\phi}^2}{M_P}
     \end{pmatrix}.
\end{align}

This is a minimal inverse seesaw structure in which lepton-number violation enters only through the gravity-induced $(3,3)$ entry. The seesaw relation then yields the light neutrino mass
\begin{align}
\label{eq:numassesEX}
   m_\nu = \mathcal{G}_i \dfrac{\mathcal{C}_1^2}{\mathcal{C}_2^2} \dfrac{v^2}{M_P} \dfrac{\Lambda^{4}}{\vev{\phi}^{4}} \, .
\end{align}

Compared to the Planck-suppressed Weinberg operator (see Eq.~\eqref{eq:Weinbergdirect}), neutrino masses in this scenario receive a large enhancement proportional to $\varepsilon^{-4}$. This follows directly from Eq.~\eqref{eq:numasses}, or equivalently from the inverse seesaw structure of $\mathcal{M}_\nu$. Although possible, it is a non-trivial exercise to recover Eq.~\eqref{eq:numassesEX} from a standard Effective Field Theory approach, since the leading operator contributing to neutrino masses is high dimensional and would naively be expected to be strongly suppressed by the scales of new physics, $\Lambda$ and $M_P$.

By construction, the light-neutrino masses arise only from Planck-suppressed effects, since the remnant $Z_3$ symmetry protects these to be vanishing in the limit of $M_P \to \infty$. This limit restores a lepton-number symmetry, and the smallness of $m_\nu$ is technically natural. 
Eq.~\eqref{eq:numassesEX} fixes the ratio $\vev{\phi}/\Lambda \approx \sqrt[4]{v^2/(m_\nu M_P)} \approx 8 \cdot 10^{-2}$, assuming $\mathcal{O}(1)$ values for the dimensionless parameters $\mathcal{C}_1$, $\mathcal{C}_2$ and $\mathcal{G}_i$ and $m_\nu = 0.05$ eV. 
We can also compute $M_R$ and $\Theta$ as
\begin{subequations}\label{eq:MRTheta}
\begin{align} 
    M_R &\approx \mathcal{C}_2 \dfrac{\langle \phi \rangle^8}{ \Lambda^7} = \mathcal{C}_2 \left( \dfrac{\mathcal{G}_i \mathcal{C}_1^2}{\mathcal{C}_2^2} \dfrac{v^2}{m_\nu M_P} \right)^{7/4}  \, \vev{\phi} \,, \\
    \Theta &\approx \dfrac{\mathcal{C}_1 \, v \varepsilon^{n_1}}{\mathcal{C}_2 \langle \phi \rangle \varepsilon^{n_2}} = \dfrac{\mathcal{C}_1 v}{M_R} \left(\dfrac{\mathcal{G}_i \mathcal{C}_1^2}{\mathcal{C}_2^2}\dfrac{ v^2 }{ m_\nu M_P} \right)^{5/4} \, .
\end{align}
\end{subequations}

Thus, as mentioned above, the entire seesaw sector is controlled by a single scale $\vev{\phi}$. While we have focused on a specific choice of Eq.~\eqref{eq:n1n2FCC}, this feature persists in more general constructions with different intermediate-sector charges. In those cases the functional dependence of $M_R$ and $\Theta$ on $\vev{\phi}$ is preserved, but the powers of $m_\nu$ and $M_P$ appearing in Eq.~\eqref{eq:MRTheta} are modified. Across this broader class of gravity-assisted low-scale seesaws, the symmetry-breaking scale continues to set the overall size of the seesaw and its phenomenology, while different intermediate-sector structures can significantly alter the predicted pattern of heavy-neutrino masses and mixings.

\begin{figure}
    \centering
    \includegraphics[width=0.9 \linewidth]{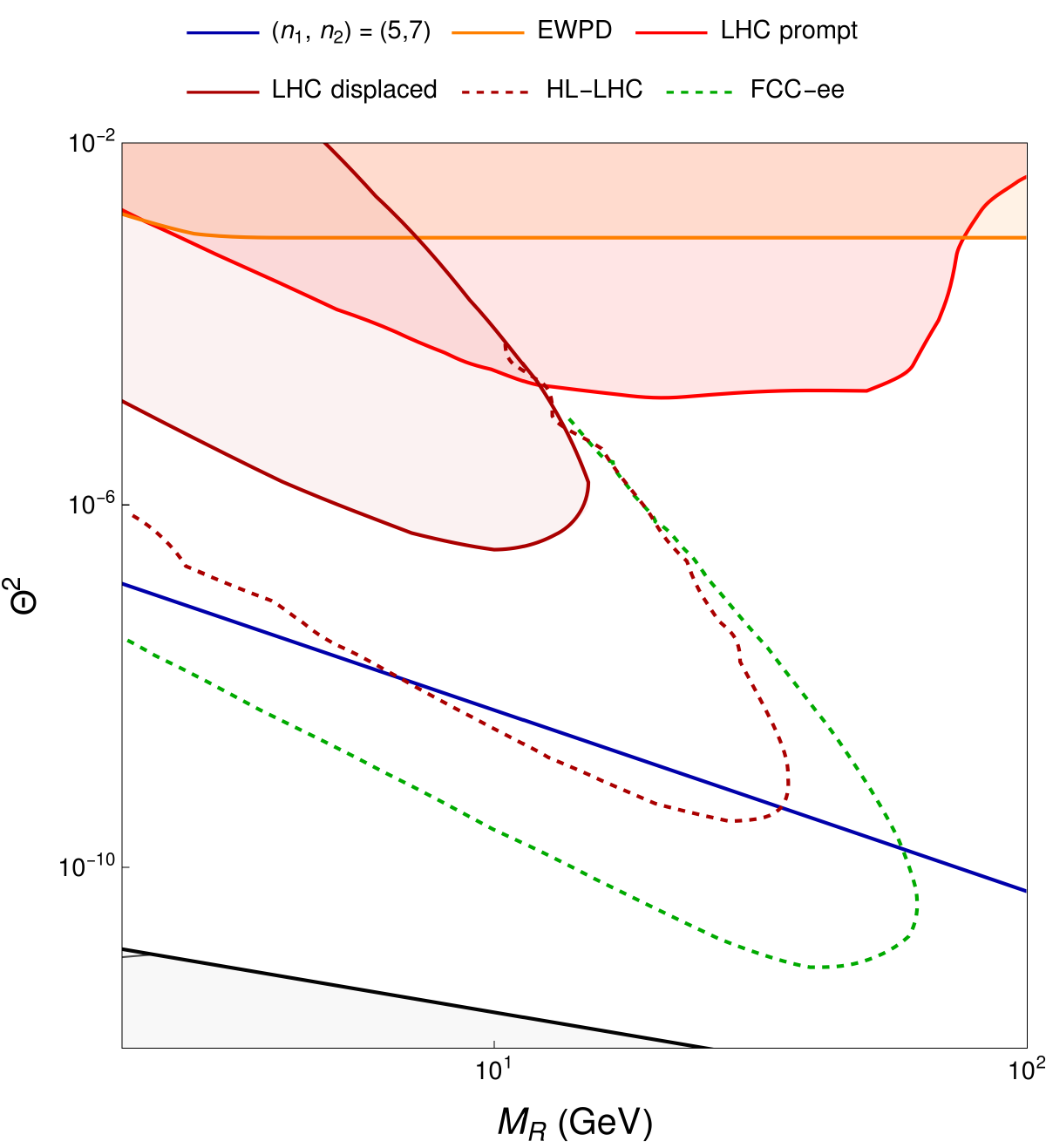}
    \caption{Predictions for $\Theta^2(M_R)$ in the case of $(n_1, n_2)=(5,7)$, shown in Eq.~\eqref{eq:MRTheta}, with coefficients set to one. The shaded region on the bottom is the area below the conventional seesaw line $\Theta^2 = m_\nu/M_R$. 
    In solid, we show the current experimental bounds of LHC prompt searches~\cite{CMS:2018iaf} (red), LHC displaced vertex searches~\cite{CMS:2022fut} (dark red), electroweak precision data~\cite{delAguila:2008pw, deBlas:2013gla, Antusch:2014woa, Blennow:2016jkn} (orange), and in dashed the future sensitivities of HL-LHC displaced vertex searches~\cite{Drewes:2019fou} (dark red), and of the FCC-ee~\cite{Blondel:2022qqo} (green), with most of the data taken from~\cite{Bolton:2019pcu}.}  
    \label{fig:FCC}
\end{figure}

These results are obtained under the consistency conditions of sub-leading gravity-induced terms and a viable seesaw expansion (see Eqs~\eqref{eq:cond1}~and~\eqref{eq:cond2}), which hold if 
\begin{subequations}
\begin{align}
\vev{\phi} &\ll M_P \left( \dfrac{\mathcal{C}_1^2 \mathcal{G}_i}{\mathcal{C}_2^2} \dfrac{v^2}{M_P m_\nu} \right)^{7/4} \approx 3 \cdot 10^{8} \, \text{ TeV} \, ,    \\ 
\vev{\phi} &\gg  \dfrac{\mathcal{C}_2 }{\mathcal{C}_1} \sqrt{\dfrac{m_\nu M_P}{\mathcal{G}_i}}   \approx 24 \,  \text{ TeV} \, .
\end{align}
\end{subequations}
These bounds define the region of parameter space allowed in this model for $M_R$ and $\Theta$, up to $\mathcal{O}(1)$ parameters.
In particular, we find the upper and lower bounds for $M_R$ and $\Theta$ as:
\begin{align}
1 \text{ keV} \ll M_R \ll 11 \text{ TeV} \, ,    \qquad   
1 \gg \Theta \gg  7 \times 10^{-8}  \, .\label{eq:predictions}
\end{align}
For coefficients set to one, the model predicts a line in the $(M_R,\Theta^2)$ plane. While beam dump experiments exclude the region of {$M_R \lesssim 2$ GeV~\cite{Barouki:2022bkt}}, the model prediction for larger masses is currently untested.
Part of it lies within the reach of future experiments such as the HL-LHC \cite{Apollinari:2017lan} and the FCC-ee~\cite{FCC:2018evy}, up to $M_R \approx 80~\text{GeV}$, as illustrated in Fig.~\ref{fig:FCC}. 

%%%%%%%%%%%%%%%%%%%%%%%%%%%%%%%
\section{Conclusions and Outlook}
%%%%%%%%%%%%%%%%%%%%%%%%%%%%%%%

In this Letter, we have introduced a novel model-building framework in which the observed smallness of light neutrino masses arises ``gravity-assisted''. While neutrino masses generated by a Planck-suppressed Weinberg operator are expected to be too small, and the standard type-I seesaw mechanism with order one couplings points to high-scale neutrino mass generation far above observational capabilities, the ``gravity-assisted'' scheme opens up new possibilities for constructing ``low-scale'' neutrino mass generation models, where light neutrino masses are ``symmetry protected'' and therefore technically natural. No small model parameters or cancellations are required in order to arrive at models that explain the small light neutrino masses while being within reach of the (HL-)LHC and future colliders and testable by precision experiments (\textit{e.g.}~on cLFV).

Gravity-assisted neutrino masses arise in the presence of a global symmetry that is spontaneously broken at an intermediate scale between the electroweak and Planck scales, leaving a remnant discrete symmetry that forbids light neutrino masses. After spontaneous symmetry breaking, masses and mixing with active neutrinos are generated for the HNLs, potentially suppressed by powers of $\vev{\phi}/\Lambda$. Nevertheless, light neutrinos remain massless due to the residual discrete symmetry, and pairs of right-chiral neutrinos form (at this level exact) heavy Dirac states. Light neutrino masses finally arise through gravity, when the residual symmetry is broken by Planck-suppressed effective operators. The result is small light neutrino masses and pseudo-Dirac pairs of heavy neutrinos with masses that can lie within the reach of future collider experiments, far below the usually expected high seesaw scale.

We described the proposed new scheme explicitly for the case of 2 HNLs and a global U(1) symmetry broken spontaneously to a $Z_3$ subgroup. It leads to a plethora of new models with expected rich phenomenology that depends on the symmetry assignments. We illustrated ``gravity-assisted'' neutrino mass generation and its phenomenological implications for collider searches in a simple benchmark realisation, leaving a more comprehensive exploration of the model class and its extensions for future work. In addition to predictions for future collider searches and precision tests such as cLFV, which may allow to test and distinguish specific model realisations, various cosmological consequences also offer interesting discovery potential. For example, the breaking of the U(1) symmetry, spontaneously as well as explicitly by gravity, leads to a Pseudo-Nambu Goldstone boson with mass linked to $M_R$. This breaking may additionally induce networks of hybrid topological defects composed of cosmic strings and domain walls, which may generate a testable contribution to the stochastic gravitational wave background.  

Beyond the class of models described above, various extensions, generalisations and embeddings into SM extensions towards solving other challenges in particle physics and cosmology appear interesting subjects for future investigations. For example, different choices of $G_\Lambda$ and $G_{\rm res}$ may be used as an alternative to $U(1)$ and $Z_3$. $G_\Lambda$ and its charge assignments could be linked to explanations of the flavour puzzle, in particular to the generation of the charged fermion mass hierarchy and large neutrino mixing. Further in this direction, embedding of the framework into Grand Unified Theories and its supersymmetric realisation have the potential to provide additional testable signatures. In cosmology, it will be interesting to study (low-scale)  leptogenesis, as well as possible links to dark matter and inflation. Finally, the idea may also be applied to other low-scale mechanisms of neutrino mass generation. In summary, the approach of ``gravity-assisted'' neutrino mass generation opens up various new possibilities for constructing testable models explaining the origin of the observed light neutrino masses.

\section{Acknowledgements}

This work was supported by the program ``Swiss High Energy Physics for the FCC'' (CHEF). SCCh acknowledges support from the Spanish grants PID2023-147306NB-I00, CNS2024-154524 and CEX2023-001292-S (MICIU/AEI/10.13039/501100011033). SCCh would like to thank the hospitality of University of Basel, where this project was initiated. 

\bibliography{biby.bib}

@article{Apollinari:2017lan,
    editor = {Apollinari, G. and B{\'e}jar Alonso, I. and Br{\"u}ning, O. and Fessia, P. and Lamont, M. and Rossi, L. and Tavian, L.},
    title = "{High-Luminosity Large Hadron Collider (HL-LHC)}: {Technical Design Report V. 0.1}",
    reportNumber = "CERN-2017-007-M",
    doi = "10.23731/CYRM-2017-004",
    volume = "4/2017",
    year = "2017"
}

@article{Antusch:2020pnn,
    author = "Antusch, Stefan and Rosskopp, Johannes",
    title = "{Heavy Neutrino-Antineutrino Oscillations in Quantum Field Theory}",
    eprint = "2012.05763",
    archivePrefix = "arXiv",
    primaryClass = "hep-ph",
    doi = "10.1007/JHEP03(2021)170",
    journal = "JHEP",
    volume = "03",
    pages = "170",
    year = "2021"
}

@article{Antusch:2006vwa,
    author = "Antusch, S. and Biggio, C. and Fernandez-Martinez, E. and Gavela, M. B. and Lopez-Pavon, J.",
    title = "{Unitarity of the Leptonic Mixing Matrix}",
    eprint = "hep-ph/0607020",
    archivePrefix = "arXiv",
    reportNumber = "FTUAM-06-8, IFT-UAM-CSIC-06-30",
    doi = "10.1088/1126-6708/2006/10/084",
    journal = "JHEP",
    volume = "10",
    pages = "084",
    year = "2006"
}

@article{Antusch:2024otj,
    author = "Antusch, Stefan and Hajer, Jan and Oliveira, Bruno M. S.",
    title = "{Discovering heavy neutrino-antineutrino oscillations at the Z-pole}",
    eprint = "2408.01389",
    archivePrefix = "arXiv",
    primaryClass = "hep-ph",
    doi = "10.1007/JHEP11(2024)102",
    journal = "JHEP",
    volume = "11",
    pages = "102",
    year = "2024"
}

@article{Antusch:2023nqd,
    author = "Antusch, Stefan and Hajer, Jan and Rosskopp, Johannes",
    title = "{Decoherence effects on lepton number violation from heavy neutrino-antineutrino oscillations}",
    eprint = "2307.06208",
    archivePrefix = "arXiv",
    primaryClass = "hep-ph",
    doi = "10.1007/JHEP11(2023)235",
    journal = "JHEP",
    volume = "11",
    pages = "235",
    year = "2023"
}

@article{Antusch:2022hhh,
    author = "Antusch, Stefan and Hajer, Jan and Rosskopp, Johannes",
    title = "{Beyond lepton number violation at the HL-LHC: resolving heavy neutrino-antineutrino oscillations}",
    eprint = "2212.00562",
    archivePrefix = "arXiv",
    primaryClass = "hep-ph",
    doi = "10.1007/JHEP09(2023)170",
    journal = "JHEP",
    volume = "09",
    pages = "170",
    year = "2023"
}

@article{Antusch:2022ceb,
    author = "Antusch, Stefan and Hajer, Jan and Rosskopp, Johannes",
    title = "{Simulating lepton number violation induced by heavy neutrino-antineutrino oscillations at colliders}",
    eprint = "2210.10738",
    archivePrefix = "arXiv",
    primaryClass = "hep-ph",
    doi = "10.1007/JHEP03(2023)110",
    journal = "JHEP",
    volume = "03",
    pages = "110",
    year = "2023"
}

@article{Antusch:2019eiz,
    author = "Antusch, Stefan and Fischer, Oliver and Hammad, A.",
    title = "{Lepton-Trijet and Displaced Vertex Searches for Heavy Neutrinos at Future Electron-Proton Colliders}",
    eprint = "1908.02852",
    archivePrefix = "arXiv",
    primaryClass = "hep-ph",
    doi = "10.1007/JHEP03(2020)110",
    journal = "JHEP",
    volume = "03",
    pages = "110",
    year = "2020"
}

@article{Antusch:2018bgr,
    author = "Antusch, Stefan and Cazzato, Eros and Fischer, Oliver and Hammad, A. and Wang, Kechen",
    title = "{Lepton Flavor Violating Dilepton Dijet Signatures from Sterile Neutrinos at Proton Colliders}",
    eprint = "1805.11400",
    archivePrefix = "arXiv",
    primaryClass = "hep-ph",
    reportNumber = "DESY 17-151, DESY-17-151",
    doi = "10.1007/JHEP10(2018)067",
    journal = "JHEP",
    volume = "10",
    pages = "067",
    year = "2018"
}

@article{Antusch:2017pkq,
    author = "Antusch, Stefan and Cazzato, Eros and Drewes, Marco and Fischer, Oliver and Garbrecht, Bjorn and Gueter, Dario and Klaric, Juraj",
    title = "{Probing Leptogenesis at Future Colliders}",
    eprint = "1710.03744",
    archivePrefix = "arXiv",
    primaryClass = "hep-ph",
    reportNumber = "TUM-1160/18, CP3-17-48",
    doi = "10.1007/JHEP09(2018)124",
    journal = "JHEP",
    volume = "09",
    pages = "124",
    year = "2018"
}

@article{Antusch:2017ebe,
    author = "Antusch, Stefan and Cazzato, Eros and Fischer, Oliver",
    title = "{Resolvable heavy neutrino{\textendash}antineutrino oscillations at colliders}",
    eprint = "1709.03797",
    archivePrefix = "arXiv",
    primaryClass = "hep-ph",
    doi = "10.1142/S0217732319500615",
    journal = "Mod. Phys. Lett. A",
    volume = "34",
    number = "07n08",
    pages = "1950061",
    year = "2019"
}

@article{Antusch:2016qby,
    author = "Antusch, Stefan and Cazzato, Eros and Fischer, Oliver",
    editor = "Flores Castillo, L. R. and Prokofiev, K.",
    title = "{Higgs production through sterile neutrinos}",
    doi = "10.1142/S0217751X16440073",
    journal = "Int. J. Mod. Phys. A",
    volume = "31",
    number = "33",
    pages = "1644007",
    year = "2016"
}

@article{Antusch:2017hhu,
    author = "Antusch, Stefan and Cazzato, Eros and Fischer, Oliver",
    title = "{Sterile neutrino searches via displaced vertices at LHCb}",
    eprint = "1706.05990",
    archivePrefix = "arXiv",
    primaryClass = "hep-ph",
    doi = "10.1016/j.physletb.2017.09.057",
    journal = "Phys. Lett. B",
    volume = "774",
    pages = "114--118",
    year = "2017"
}

@article{Antusch:2016vyf,
    author = "Antusch, Stefan and Cazzato, Eros and Fischer, Oliver",
    title = "{Displaced vertex searches for sterile neutrinos at future lepton colliders}",
    eprint = "1604.02420",
    archivePrefix = "arXiv",
    primaryClass = "hep-ph",
    doi = "10.1007/JHEP12(2016)007",
    journal = "JHEP",
    volume = "12",
    pages = "007",
    year = "2016"
}

@article{Antusch:2008tz,
    author = "Antusch, Stefan and Baumann, Jochen P. and Fernandez-Martinez, Enrique",
    title = "{Non-Standard Neutrino Interactions with Matter from Physics Beyond the Standard Model}",
    eprint = "0807.1003",
    archivePrefix = "arXiv",
    primaryClass = "hep-ph",
    reportNumber = "MPP-2008-74",
    doi = "10.1016/j.nuclphysb.2008.11.018",
    journal = "Nucl. Phys. B",
    volume = "810",
    pages = "369--388",
    year = "2009"
}

@article{Antusch:2017tud,
    author = "Antusch, Stefan and Hohl, Christian and King, Steve F. and Susic, Vasja",
    title = "{Non-universal Z' from SO(10) GUTs with vector-like family and the origin of neutrino masses}",
    eprint = "1712.05366",
    archivePrefix = "arXiv",
    primaryClass = "hep-ph",
    doi = "10.1016/j.nuclphysb.2018.07.022",
    journal = "Nucl. Phys. B",
    volume = "934",
    pages = "578--605",
    year = "2018"
}

@article{deGouvea:2000jp,
    author = "de Gouvea, Andre and Valle, J. W. F.",
    title = "{Minimalistic neutrino mass model}",
    eprint = "hep-ph/0010299",
    archivePrefix = "arXiv",
    reportNumber = "CERN-TH-2000-319, IFIC-00-63",
    doi = "10.1016/S0370-2693(01)00103-4",
    journal = "Phys. Lett. B",
    volume = "501",
    pages = "115--127",
    year = "2001"
}

@article{Carvajal:2015dxa,
    author = "Carvajal, C. D. R. and Dias, A. G. and Nishi, C. C. and S{\'a}nchez-Vega, B. L.",
    title = "{Axion Like Particles and the Inverse Seesaw Mechanism}",
    eprint = "1503.03502",
    archivePrefix = "arXiv",
    primaryClass = "hep-ph",
    doi = "10.1007/JHEP05(2015)069",
    journal = "JHEP",
    volume = "05",
    pages = "069",
    year = "2015",
    note = "[Erratum: JHEP 08, 103 (2015)]"
}

@article{Das:2012ze,
    author = "Das, Arindam and Okada, Nobuchika",
    title = "{Inverse seesaw neutrino signatures at the LHC and ILC}",
    eprint = "1207.3734",
    archivePrefix = "arXiv",
    primaryClass = "hep-ph",
    doi = "10.1103/PhysRevD.88.113001",
    journal = "Phys. Rev. D",
    volume = "88",
    pages = "113001",
    year = "2013"
}

@article{Antusch:2014woa,
    author = "Antusch, Stefan and Fischer, Oliver",
    title = "{Non-unitarity of the leptonic mixing matrix: Present bounds and future sensitivities}",
    eprint = "1407.6607",
    archivePrefix = "arXiv",
    primaryClass = "hep-ph",
    reportNumber = "MPP-2014-313",
    doi = "10.1007/JHEP10(2014)094",
    journal = "JHEP",
    volume = "10",
    pages = "094",
    year = "2014"
}

@article{Ilakovac:1994kj,
    author = "Ilakovac, A. and Pilaftsis, A.",
    title = "{Flavor violating charged lepton decays in seesaw-type models}",
    eprint = "hep-ph/9403398",
    archivePrefix = "arXiv",
    reportNumber = "RAL-94-032, MZ-TH-94-15",
    doi = "10.1016/0550-3213(94)00567-X",
    journal = "Nucl. Phys. B",
    volume = "437",
    pages = "491",
    year = "1995"
}

@article{delAguila:2007qnc,
    author = "del Aguila, F. and Aguilar-Saavedra, J. A. and Pittau, R.",
    title = "{Heavy neutrino signals at large hadron colliders}",
    eprint = "hep-ph/0703261",
    archivePrefix = "arXiv",
    doi = "10.1088/1126-6708/2007/10/047",
    journal = "JHEP",
    volume = "10",
    pages = "047",
    year = "2007"
}

@article{Fujihara:2005uq,
    author = "Fujihara, T. and Kang, S. K. and Kim, C. S. and Kimura, D. and Morozumi, T.",
    title = "{Low scale seesaw model and lepton flavor violating rare B decays}",
    eprint = "hep-ph/0512010",
    archivePrefix = "arXiv",
    reportNumber = "HUPD-0507",
    doi = "10.1103/PhysRevD.73.074011",
    journal = "Phys. Rev. D",
    volume = "73",
    pages = "074011",
    year = "2006"
}

@article{Abada:2018sfh,
      author         = "Abada, Asmaa and Bernal, Nicolás and Losada, Marta and Marcano, Xabier",
      title          = "{Inclusive displaced vertex searches for heavy neutral leptons at the LHC}",
      journal        = "JHEP",
      volume         = "01",
      year           = "2019",
      pages          = "093",
      doi            = "10.1007/JHEP01(2019)093",
      eprint         = "1807.10024",
      archivePrefix  = "arXiv",
      primaryClass   = "hep-ph",
}

@article{Bondarenko:2018ptm,
      author         = "Bondarenko, Kyrylo and Boyarsky, Alexey and Gorbunov, Dmitry and Ruchayskiy, Oleg",
      title          = "{Phenomenology of GeV-scale Heavy Neutral Leptons}",
      journal        = "JHEP",
      volume         = "11",
      year           = "2018",
      pages          = "032",
      doi            = "10.1007/JHEP11(2018)032",
      eprint         = "1805.08567",
      archivePrefix  = "arXiv",
      primaryClass   = "hep-ph",
}

@article{Antusch:2016ejd,
      author         = "Antusch, Stefan and Cazzato, Eros and Fischer, Oliver",
      title          = "{Displaced vertex searches for sterile neutrinos at future lepton colliders}",
      journal        = "JHEP",
      volume         = "12",
      year           = "2016",
      pages          = "007",
      doi            = "10.1007/JHEP12(2016)007",
      eprint         = "1604.02420",
      archivePrefix  = "arXiv",
      primaryClass   = "hep-ph",
}

@article{Borah:2019ldn,
    author = "Borah, Debasish and Karmakar, Biswajit and Nanda, Dibyendu",
    title = "{Planck scale origin of nonzero $\theta_{13}$ and super-WIMP dark matter}",
    eprint = "1906.02756",
    archivePrefix = "arXiv",
    primaryClass = "hep-ph",
    doi = "10.1103/PhysRevD.100.055014",
    journal = "Phys. Rev. D",
    volume = "100",
    number = "5",
    pages = "055014",
    year = "2019"
}

@article{Borah:2013mqa,
    author = "Borah, Debasish",
    title = "{Effects of Planck Scale Physics on Neutrino Mixing Parameters in Left-Right Symmetric Models}",
    eprint = "1305.1254",
    archivePrefix = "arXiv",
    primaryClass = "hep-ph",
    doi = "10.1103/PhysRevD.87.095009",
    journal = "Phys. Rev. D",
    volume = "87",
    number = "9",
    pages = "095009",
    year = "2013"
}

@article{Senjanovic:2020rcq,
    author = "Senjanovic, Goran",
    title = "{Neutrino 2020: Theory Outlook}",
    eprint = "2011.01264",
    archivePrefix = "arXiv",
    primaryClass = "hep-ph",
    doi = "10.1142/S0217751X21300039",
    journal = "Int. J. Mod. Phys. A",
    volume = "36",
    number = "02",
    pages = "2130003",
    year = "2021"
}

@article{Borah:2020ljr,
    author = "Borah, Debasish and Jyoti Das, Suruj and Saha, Abhijit Kumar",
    title = "{Gravitational origin of dark matter and Majorana neutrino mass with non-minimal quartic inflation}",
    eprint = "2011.02489",
    archivePrefix = "arXiv",
    primaryClass = "hep-ph",
    doi = "10.1016/j.dark.2021.100858",
    journal = "Phys. Dark Univ.",
    volume = "33",
    pages = "100858",
    year = "2021"
}

@article{Barenboim:2019fmj,
    author = "Barenboim, Gabriela and Turner, Jessica and Zhou, Ye-Ling",
    title = "{Light neutrino masses from gravitational condensation: the Schwinger{\textendash}Dyson approach}",
    eprint = "1909.04675",
    archivePrefix = "arXiv",
    primaryClass = "hep-ph",
    reportNumber = "FERMILAB-PUB-19-461-T",
    doi = "10.1140/epjc/s10052-021-09300-8",
    journal = "Eur. Phys. J. C",
    volume = "81",
    number = "6",
    pages = "511",
    year = "2021"
}

@article{Ibarra:2018dib,
    author = "Ibarra, Alejandro and Strobl, Patrick and Toma, Takashi",
    title = "{Neutrino masses from Planck-scale lepton number breaking}",
    eprint = "1802.09997",
    archivePrefix = "arXiv",
    primaryClass = "hep-ph",
    reportNumber = "TUM-HEP-1133-18, KIAS-P18015",
    doi = "10.1103/PhysRevLett.122.081803",
    journal = "Phys. Rev. Lett.",
    volume = "122",
    number = "8",
    pages = "081803",
    year = "2019"
}

@article{Bolton:2019pcu,
    author = "Bolton, Patrick D. and Deppisch, Frank F. and Bhupal Dev, P. S.",
    title = "{Neutrinoless double beta decay versus other probes of heavy sterile neutrinos}",
    eprint = "1912.03058",
    archivePrefix = "arXiv",
    primaryClass = "hep-ph",
    doi = "10.1007/JHEP03(2020)170",
    journal = "JHEP",
    volume = "03",
    pages = "170",
    year = "2020"
}

@article{Barouki:2022bkt,
    author = "Barouki, Ryan and Marocco, Giacomo and Sarkar, Subir",
    title = "{Blast from the past II: Constraints on heavy neutral leptons from the BEBC WA66 beam dump experiment}",
    eprint = "2208.00416",
    archivePrefix = "arXiv",
    primaryClass = "hep-ph",
    doi = "10.21468/SciPostPhys.13.5.118",
    journal = "SciPost Phys.",
    volume = "13",
    pages = "118",
    year = "2022"
}

@article{delAguila:2008pw,
    author = "del Aguila, F. and de Blas, J. and Perez-Victoria, M.",
    title = "{Effects of new leptons in Electroweak Precision Data}",
    eprint = "0803.4008",
    archivePrefix = "arXiv",
    primaryClass = "hep-ph",
    reportNumber = "UG-FT-224-08, CAFPE-94-08",
    doi = "10.1103/PhysRevD.78.013010",
    journal = "Phys. Rev. D",
    volume = "78",
    pages = "013010",
    year = "2008"
}

@article{deBlas:2013gla,
    author = "de Blas, J.",
    editor = "Bosman, M. and Juste, A. and Mart{\'\i}nez, M. and Sorin, V.",
    title = "{Electroweak limits on physics beyond the Standard Model}",
    eprint = "1307.6173",
    archivePrefix = "arXiv",
    primaryClass = "hep-ph",
    doi = "10.1051/epjconf/20136019008",
    journal = "EPJ Web Conf.",
    volume = "60",
    pages = "19008",
    year = "2013"
}

@article{Blennow:2016jkn,
    author = "Blennow, Mattias and Coloma, Pilar and Fernandez-Martinez, Enrique and Hernandez-Garcia, Josu and Lopez-Pavon, Jacobo",
    title = "{Non-Unitarity, sterile neutrinos, and Non-Standard neutrino Interactions}",
    eprint = "1609.08637",
    archivePrefix = "arXiv",
    primaryClass = "hep-ph",
    reportNumber = "IFT-UAM-CSIC-16-090, FTUAM-16-35, FERMILAB-PUB-16-400-T",
    doi = "10.1007/JHEP04(2017)153",
    journal = "JHEP",
    volume = "04",
    pages = "153",
    year = "2017"
}

@article{CMS:2022fut,
    author = "Tumasyan, Armen and others",
    collaboration = "CMS",
    title = "{Search for long-lived heavy neutral leptons with displaced vertices in proton-proton collisions at $ \sqrt{\mathrm{s}} $ =13 TeV}",
    eprint = "2201.05578",
    archivePrefix = "arXiv",
    primaryClass = "hep-ex",
    reportNumber = "CMS-EXO-20-009, CERN-EP-2021-264",
    doi = "10.1007/JHEP07(2022)081",
    journal = "JHEP",
    volume = "07",
    pages = "081",
    year = "2022"
}

@article{Blondel:2022qqo,
    author = "Blondel, A. and others",
    title = "{Searches for long-lived particles at the future FCC-ee}",
    eprint = "2203.05502",
    archivePrefix = "arXiv",
    primaryClass = "hep-ex",
    doi = "10.3389/fphy.2022.967881",
    journal = "Front. in Phys.",
    volume = "10",
    pages = "967881",
    year = "2022"
}

@article{Dvali:2016uhn,
    author = "Dvali, Gia and Funcke, Lena",
    title = "{Small neutrino masses from gravitational {\ensuremath{\theta}}-term}",
    eprint = "1602.03191",
    archivePrefix = "arXiv",
    primaryClass = "hep-ph",
    reportNumber = "MPP-2016-277, LMU-ASC-31-16",
    doi = "10.1103/PhysRevD.93.113002",
    journal = "Phys. Rev. D",
    volume = "93",
    number = "11",
    pages = "113002",
    year = "2016"
}

@article{Davoudiasl:2020opf,
    author = "Davoudiasl, Hooman",
    title = "{Gravitational interactions and neutrino masses}",
    eprint = "2003.04908",
    archivePrefix = "arXiv",
    primaryClass = "hep-ph",
    doi = "10.1103/PhysRevD.101.115024",
    journal = "Phys. Rev. D",
    volume = "101",
    number = "11",
    pages = "115024",
    year = "2020"
}

@article{Grasso:1992fv,
    author = "Grasso, D. and Lusignoli, Maurizio and Roncadelli, M.",
    title = "{Global continuous symmetry and the 17-keV neutrino}",
    reportNumber = "ROME-868-1992, FNT-T-92-07",
    doi = "10.1016/0370-2693(92)91967-E",
    journal = "Phys. Lett. B",
    volume = "288",
    pages = "140--144",
    year = "1992"
}

@article{Akhmedov:1992hh,
    author = "Akhmedov, Evgeny K. and Berezhiani, Zurab G. and Senjanovic, Goran",
    title = "{Planck scale physics and neutrino masses}",
    eprint = "hep-ph/9205230",
    archivePrefix = "arXiv",
    reportNumber = "SISSA-83-92-EP, LMU-04-92, IC-92-79",
    doi = "10.1103/PhysRevLett.69.3013",
    journal = "Phys. Rev. Lett.",
    volume = "69",
    pages = "3013--3016",
    year = "1992"
}

@article{Banks:1988yz,
    author = "Banks, Tom and Dixon, Lance J.",
    title = "{Constraints on String Vacua with Space-Time Supersymmetry}",
    reportNumber = "PUPT-1086, SCIPP-8805",
    doi = "10.1016/0550-3213(88)90523-8",
    journal = "Nucl. Phys. B",
    volume = "307",
    pages = "93--108",
    year = "1988"
}

@article{Banks:2010zn,
    author = "Banks, Tom and Seiberg, Nathan",
    title = "{Symmetries and Strings in Field Theory and Gravity}",
    eprint = "1011.5120",
    archivePrefix = "arXiv",
    primaryClass = "hep-th",
    doi = "10.1103/PhysRevD.83.084019",
    journal = "Phys. Rev. D",
    volume = "83",
    pages = "084019",
    year = "2011"
}

@article{Harlow:2018tng,
    author = "Harlow, Daniel and Ooguri, Hirosi",
    title = "{Symmetries in quantum field theory and quantum gravity}",
    eprint = "1810.05338",
    archivePrefix = "arXiv",
    primaryClass = "hep-th",
    doi = "10.1007/s00220-021-04040-y",
    journal = "Commun. Math. Phys.",
    volume = "383",
    number = "3",
    pages = "1669--1804",
    year = "2021"
}

@article{Barbieri:1979hc,
    author = "Barbieri, Riccardo and Ellis, John R. and Gaillard, Mary K.",
    title = "{Neutrino Masses and Oscillations in SU(5)}",
    reportNumber = "CERN-TH-2787, LAPP-TH-10",
    doi = "10.1016/0370-2693(80)90734-0",
    journal = "Phys. Lett. B",
    volume = "90",
    pages = "249--252",
    year = "1980"
}

@article{Hawking:1975vcx,
    author = "Hawking, S. W.",
    editor = "Gibbons, G. W. and Hawking, S. W.",
    title = "{Particle Creation by Black Holes}",
    doi = "10.1007/BF02345020",
    journal = "Commun. Math. Phys.",
    volume = "43",
    pages = "199--220",
    year = "1975",
    note = "[Erratum: Commun.Math.Phys. 46, 206 (1976)]"
}

@article{Kallosh:1995hi,
    author = "Kallosh, Renata and Linde, Andrei D. and Linde, Dmitri A. and Susskind, Leonard",
    title = "{Gravity and global symmetries}",
    eprint = "hep-th/9502069",
    archivePrefix = "arXiv",
    reportNumber = "SU-ITP-95-2",
    doi = "10.1103/PhysRevD.52.912",
    journal = "Phys. Rev. D",
    volume = "52",
    pages = "912--935",
    year = "1995"
}

@article{deSalas:2020pgw,
    author = "de Salas, P. F. and Forero, D. V. and Gariazzo, S. and Mart{\'\i}nez-Mirav{\'e}, P. and Mena, O. and Ternes, C. A. and T{\'o}rtola, M. and Valle, J. W. F.",
    title = "{2020 global reassessment of the neutrino oscillation picture}",
    eprint = "2006.11237",
    archivePrefix = "arXiv",
    primaryClass = "hep-ph",
    doi = "10.1007/JHEP02(2021)071",
    journal = "JHEP",
    volume = "02",
    pages = "071",
    year = "2021"
}

@article{CentellesChulia:2024uzv,
    author = "Centelles Chuli\'a, Salvador and Herrero-Brocal, Antonio and Vicente, Avelino",
    title = "{The Type-I Seesaw family}",
    eprint = "2404.15415",
    archivePrefix = "arXiv",
    primaryClass = "hep-ph",
    month = "4",
    year = "2024"
}

@article{Minkowski:1977sc,
    author = "Minkowski, Peter",
    title = "{$\mu \to e\gamma$ at a Rate of One Out of $10^{9}$ Muon Decays?}",
    reportNumber = "Print-77-0182 (BERN)",
    doi = "10.1016/0370-2693(77)90435-X",
    journal = "Phys. Lett. B",
    volume = "67",
    pages = "421--428",
    year = "1977"
}

@Article{FCC:2018evy,
        author = "Abada, A. and others",
 collaboration = "FCC",
         title = "{FCC-ee: The Lepton Collider}",
       journal = "Eur.Phys.J.ST",
        volume = "228",
          year = "2019",
         pages = "261-623",
           doi = "10.1140/epjst/e2019-900045-4",
        number = "2",
  reportnumber = "CERN-ACC-2018-0057",
}

@article{Hagedorn:2021ldq,
    author = "Hagedorn, C. and Kriewald, J. and Orloff, J. and Teixeira, A. M.",
    title = "{Flavour and CP symmetries in the inverse seesaw}",
    eprint = "2107.07537",
    archivePrefix = "arXiv",
    primaryClass = "hep-ph",
    reportNumber = "IFIC/21-21, FTUV-21-0611.3946",
    doi = "10.1140/epjc/s10052-022-10097-3",
    journal = "Eur. Phys. J. C",
    volume = "82",
    number = "3",
    pages = "194",
    year = "2022"
}

@article{Hirsch:2020klk,
    author = "Hirsch, Martin and Wang, Zeren Simon",
    title = "{Heavy neutral leptons at ANUBIS}",
    eprint = "2001.04750",
    archivePrefix = "arXiv",
    primaryClass = "hep-ph",
    reportNumber = "APCTP Pre2020-002, IFIC/20-01",
    doi = "10.1103/PhysRevD.101.055034",
    journal = "Phys. Rev. D",
    volume = "101",
    number = "5",
    pages = "055034",
    year = "2020"
}

@article{Abada:2014kba,
    author = "Abada, A. and Krauss, Manuel E. and Porod, W. and Staub, F. and Vicente, A. and Weiland, Cedric",
    title = "{Lepton flavor violation in low-scale seesaw models: SUSY and non-SUSY contributions}",
    eprint = "1408.0138",
    archivePrefix = "arXiv",
    primaryClass = "hep-ph",
    reportNumber = "LPT-ORSAY-14-43, BONN-TH-14-11, IFT-UAM-CSIC-14-061, FTUAM-14-25",
    doi = "10.1007/JHEP11(2014)048",
    journal = "JHEP",
    volume = "11",
    pages = "048",
    year = "2014"
}

@article{Atre:2009rg,
    author = "Atre, Anupama and Han, Tao and Pascoli, Silvia and Zhang, Bin",
    title = "{The Search for Heavy Majorana Neutrinos}",
    eprint = "0901.3589",
    archivePrefix = "arXiv",
    primaryClass = "hep-ph",
    reportNumber = "FERMILAB-PUB-08-086-T, NSF-KITP-08-54, MADPH-06-1466, DCPT-07-198, IPPP-07-99",
    doi = "10.1088/1126-6708/2009/05/030",
    journal = "JHEP",
    volume = "05",
    pages = "030",
    year = "2009"
}

@article{Batra:2023mds,
    author = "Batra, Aditya and Bharadwaj, Praveen and Mandal, Sanjoy and Srivastava, Rahul and Valle, Jos\'e W. F.",
    title = "{Phenomenology of the simplest linear seesaw mechanism}",
    eprint = "2305.00994",
    archivePrefix = "arXiv",
    primaryClass = "hep-ph",
    doi = "10.1007/JHEP07(2023)221",
    journal = "JHEP",
    volume = "07",
    pages = "221",
    year = "2023"
}

@article{Lindner:2016bgg,
    author = "Lindner, Manfred and Platscher, Moritz and Queiroz, Farinaldo S.",
    title = "{A Call for New Physics : The Muon Anomalous Magnetic Moment and Lepton Flavor Violation}",
    eprint = "1610.06587",
    archivePrefix = "arXiv",
    primaryClass = "hep-ph",
    doi = "10.1016/j.physrep.2017.12.001",
    journal = "Phys. Rept.",
    volume = "731",
    pages = "1--82",
    year = "2018"
}

@article{McDonald:2016ixn,
        author = "McDonald, Arthur B.",
         title = "{Nobel Lecture: The Sudbury Neutrino Observatory: Observation of flavor change for solar neutrinos}",
       journal = "Rev.Mod.Phys.",
        volume = "88",
          year = "2016",
         pages = "030502",
           doi = "10.1103/RevModPhys.88.030502",
}

@article{KamLAND:2002uet,
        author = "Eguchi, K. and others",
 collaboration = "KamLAND",
         title = "{First results from KamLAND: Evidence for reactor anti-neutrino disappearance}",
       journal = "Phys.Rev.Lett.",
        volume = "90",
          year = "2003",
         pages = "021802",
 archiveprefix = "arXiv",
  primaryclass = "hep-ex",
        eprint = "hep-ex/0212021",
           doi = "10.1103/PhysRevLett.90.021802",
  reportnumber = "hep-ex/0212021",
}

@article{GonzalezGarcia:1988rw,
        author = "Gonzalez-Garcia, M.C. and Valle, J.~W.~F.",
         title = "{Fast Decaying Neutrinos and Observable Flavor Violation in a New Class of Majoron Models}",
       journal = "Phys.Lett.",
        volume = "B216",
          year = "1989",
         pages = "360-366",
           doi = "10.1016/0370-2693(89)91131-3",
  reportnumber = "FTUV-10-88",
}

@Inproceedings{Mohapatra:1986bd,
        author = "Mohapatra, R.N. and Valle, J.~W.~F.",
         title = "{Neutrino Mass and Baryon Number Nonconservation in Superstring Models}",
       journal = "Phys.Rev.D",
        volume = "34",
          year = "1986",
         pages = "1642",
           doi = "10.1103/PhysRevD.34.1642",
  reportnumber = "MdDP-PP-86-127",
}

@article{Akhmedov:1995ip,
        author = "Akhmedov, Evgeny K. and others",
         title = "{Left-right symmetry breaking in NJL approach}",
       journal = "Phys.Lett.B",
        volume = "368",
          year = "1996",
         pages = "270-280",
 archiveprefix = "arXiv",
  primaryclass = "hep-ph",
        eprint = "hep-ph/9507275",
           doi = "10.1016/0370-2693(95)01504-3",
  reportnumber = "IC-95-125, TUM-HEP-221-95, MPI-PHT-95-35, FTUV-95-34, IFIC-95-36",
}

@article{Akhmedov:1995vm,
        author = "Akhmedov, Evgeny K. and others",
         title = "{Dynamical left-right symmetry breaking}",
       journal = "Phys.Rev.D",
        volume = "53",
          year = "1996",
         pages = "2752-2780",
 archiveprefix = "arXiv",
  primaryclass = "hep-ph",
        eprint = "hep-ph/9509255",
           doi = "10.1103/PhysRevD.53.2752",
  reportnumber = "IC-95-126, TUM-HEP-222-95, MPI-PHT-95-70, FTUV-95-36, IFIC-95-38",
}

@article{Malinsky:2005bi,
        author = "Malinsky, Michal and Romao, J.C. and Valle, J.~W.~F.",
         title = "{Novel supersymmetric SO(10) seesaw mechanism}",
       journal = "Phys.Rev.Lett.",
        volume = "95",
          year = "2005",
         pages = "161801",
 archiveprefix = "arXiv",
  primaryclass = "hep-ph",
        eprint = "hep-ph/0506296",
           doi = "10.1103/PhysRevLett.95.161801",
  reportnumber = "IFIC-05-28",
}

@article{Gonzalez-Garcia:1990sbd,
        author = "Gonzalez-Garcia, M.C. and Santamaria, A. and Valle, J.~W.~F.",
         title = "{Isosinglet Neutral Heavy Lepton Production in $Z$ Decays and Neutrino Mass}",
       journal = "Nucl.Phys.B",
        volume = "342",
          year = "1990",
         pages = "108-126",
           doi = "10.1016/0550-3213(90)90573-V",
  reportnumber = "FTUV/89-47, IFIC/89-24, MPI-PAE/PTh-2/90",
}

@article{Das:2012ii,
        author = "Das, S.P. and Deppisch, F.F. and Kittel, O. and Valle, J.~W.~F.",
         title = "{Heavy Neutrinos and Lepton Flavour Violation in Left-Right Symmetric Models at the LHC}",
       journal = "Phys.Rev.D",
        volume = "86",
          year = "2012",
         pages = "055006",
 archiveprefix = "arXiv",
  primaryclass = "hep-ph",
        eprint = "1206.0256",
           doi = "10.1103/PhysRevD.86.055006",
  reportnumber = "IFIC-12-17",
}

@article{Antusch:2015mia,
        author = "Antusch, Stefan and Fischer, Oliver",
         title = "{Testing sterile neutrino extensions of the Standard Model at future lepton colliders}",
       journal = "JHEP",
        volume = "05",
          year = "2015",
         pages = "053",
 archiveprefix = "arXiv",
  primaryclass = "hep-ph",
        eprint = "1502.05915",
           doi = "10.1007/JHEP05(2015)053",
  reportnumber = "MPP-2015-24",
}
\bibliographystyle{utphys2}

\end{document}